\documentclass[]{spie}  

\usepackage{amsmath,amsfonts,amssymb}
\usepackage{graphicx}
\usepackage{multirow,multicol}
\usepackage[font=small,labelfont=bf]{caption}
\usepackage{array}
\newcolumntype{R}[1]{>{\raggedleft\let\newline\\\arraybackslash\hspace{0pt}}m{#1}}

\title{A New Bandwidth Selection Criterion for Using SVDD to Analyze Hyperspectral Data}

\author{Yuwei Liao}
\author{Deovrat Kakde}
\author{Arin Chaudhuri}
\author{Hansi Jiang}
\author{Carol Sadek}
\author{Seunghyun Kong}

\affil{SAS Institute Inc., Cary, NC, USA}

\authorinfo{Further author information: (Send correspondence to Yuwei Liao)\\Yuwei Liao: E-mail: yuwei.liao@sas.com, Telephone: 1 919 531 2769}

\begin{document} 
\maketitle

\begin{abstract}
This paper presents a method for hyperspectral image classification that uses support vector data description (SVDD) with the Gaussian kernel function. SVDD has been a popular machine learning technique for single-class classification, but selecting the proper Gaussian kernel bandwidth to achieve the best classification performance is always a challenging problem. This paper proposes a new automatic, unsupervised Gaussian kernel bandwidth selection approach which is used with a multiclass SVDD classification scheme. The performance of the multiclass SVDD classification scheme is evaluated on three frequently used hyperspectral data sets, and preliminary results show that the proposed method can achieve better performance than published results on these data sets.  
\end{abstract}


\section{INTRODUCTION}
\label{sec:intro}  
Hyperspectral remote sensing has been an active research area for the past two decades\cite{Bioucas2013}. Varying research has been done to extract useful information from hyperspectral imaging data, which are collected from airborne or spaceborne sensors. Hypserspectral imaging data has applications in different areas such as resource management, agriculture, astronomy, mineral exploration, food inspection, and environmental monitoring\cite{Bioucas2013, Mehl2004, Feng2012, Elmasry2012, Dale2013, Wang2007, Kruse2003}. 

Identifying the contents of each pixel in 3-D hyperspectral imaging data has been a challenging problem, and various classification techniques have been studied and applied to hyperspectral data\cite{Healley1999,Walter2003}. Support vector machines (SVM) are popular classifiers because they are robust when training data samples are limited. Labeling hyperspectral data is difficult because of the complexity and heterogeneity of the geographic areas that are covered by the sensors, so the number of available labeled data samples is always limited. This limitation makes SVMs attractive in the field of hyperspectral imaging data processing\cite{Mari2007, Mari2010, Khazai2012, Dai2015}. Researchers have shown that a one-class SVM classifier can perform better than a multiclass SVM classifier if only one class is of interest in a multiclass problem\cite{Mari2007}.  

One of the well-known algorithms for one-class classification is support vector data description (SVDD), which was first introduced in Ref.~\citenum{tax2004support}.  It can be shown that SVDD formulation is equivalent to one-class SVM classification under certain conditions \cite{institute2017sas}. SVDD is used in domains where the majority of data belong to a single class, or when one of the classes is significantly undersampled. The SVDD algorithm builds a flexible boundary around the target class data; this data boundary is characterized by observations that are designated as support vectors. Applications of SVDD include machine condition monitoring \cite{widodo2007support, ypma1999robust}, image classification \cite{sanchez2007one}, and multivariate process control \cite{sukchotrat2009one, kakde2017non}. SVDD has the advantage that no assumptions about the distribution of the data need to be made. The technique can describe the shape of the target class without prior knowledge of the specific data distribution, with observations that fall outside the data boundary flagged as potential outliers. 

\section{Mathematical Formulation of SVDD}
\label{mfsvdd}
{\bf Normal Data Description}\\
The most elemental form of SVDD is a normal data description. The SVDD model for normal data description builds a minimum-radius hypersphere, which is characterized by center $a$ and radius $R(>0)$, around the data. SVDD model minimizes the volume of the sphere by minimizing $R^{2}$ and requires that the sphere contain all the training data\cite{tax2004support}. SVDD formulation can be expressed in either of the following forms:\\
\newline
{\bf Primal Form}\\

Objective function:
\begin{equation}
\min R^{2} + C\sum_{i=1}^{n}\xi _{i} 
\end{equation}
subject to: 
\begin{align}
\|x _{i}-a\|^2 \leq R^{2} + \xi_{i}, \forall i=1,\ldots,n\\
\xi _{i}\geq 0, \forall i=1,\ldots,n
\end{align}
where:\\
$x_{i} \in {\mathbb{R}}^{m}, i=1,\ldots,n  $ represents the training data,\\
$ R$ is the radius and represents the decision variable,\\
$\xi_{i}$ is the slack for each variable,\\
$a$ is the center, \\
$C=\frac{1}{nf}$ is the penalty constant that controls the tradeoff between the volume and the errors, and\\
$f$ is the expected outlier fraction.\\ \ \\
{\bf Dual Form} \label{df} \\

The dual form is obtained using the Lagrange multipliers.\\ 
Objective function:
\begin{equation} 
\max\ \sum_{i=1}^{n}\alpha _{i}(x_{i}\cdot x_{i}) - \sum_{i,j}^{ }\alpha _{i}\alpha _{j}(x_{i} \cdot x_{j}) 
\end{equation}
subject to:
\begin{align} 
&   \sum_{i=1}^{n}\alpha _{i}  = 1\\ 
&  0 \leq  \alpha_{i}\leq C,\forall i=1,\ldots,n 
\end{align}
where:\\
$\alpha_{i}\in \mathbb{R}$ are the Lagrange constants and\\
$C=\frac{1}{nf}$ is the penalty constant.\\ \ \\
{\bf Duality Information}\\

Depending upon the position of the observation, the following results hold:\\

Center position: \begin{equation} \sum_{i=1}^{n}\alpha _{i}x_{i}=a \end{equation}

Inside position: \begin{equation} \left \| x_{i}-a \right \| < R \rightarrow \alpha _{i}=0 \end{equation}

Boundary position: \begin{equation} \label{eq:9} \left \| x_{i}-a \right \| = R \rightarrow 0< \alpha _{i}< C\end{equation}

Outside position: \begin{equation} \label{eq:10} \left \| x_{i}-a \right \| > R \rightarrow \alpha _{i}= C\end{equation}
The circular data boundary can include a significant amount of space in which training observations are very sparsely distributed. Scoring with a model that has a circular data boundary can increase the probability of false positives.
Hence, instead of a circular shape, a compact bounded outline around the data is often desired. Such an outline should approximate the shape of the single-class training data and is possible with the use of kernel functions.\\
{\begin{flushleft}
		\bf {Flexible Data Description}\\	
	\end{flushleft}
	The support vector data description is made flexible by replacing the inner product $ (x_{i}\cdot x_{j}) $ with a suitable kernel function $ K(x_{i},x_{j}) $. This paper uses a Gaussian kernel function which is defined as
	\begin{equation}  \label{eq:1}
	K(x_{i}, x_{j})= \exp  \dfrac{ -\|x_i - x_j\|^2}{2s^2}
	\end{equation}
	where $s$ is the Gaussian bandwidth parameter.
	Results 7 through 10 hold when the kernel function is used in the mathematical formulation.\\
	
	The threshold $R^{2}$ is calculated as
	\begin{equation}
	R^{2} = K(x_{k},x_{k})-2\sum_{i}^{ }\alpha _{i}K(x_{i},x_{k})+\sum_{i,j}^{ }\alpha _{i}\alpha _{j}K(x_{i},x_{j})
	\end{equation}
	using any $ x_{k} \in SV_{<C} $
	, where $SV_{<C}$  is the set of support vectors for which $ \alpha _{k} < C $.
	\begin{flushleft}
		{\bf Scoring}	
	\end{flushleft}
	For each observation $ z $  in the scoring data set, the distance$\textup{ dist}^{2}(z) $ is calculated as follows: 
	\begin{equation} \textup{dist}^{2}(z)= K(z,z) - 2\sum_{i}^{ }\alpha _{i}K(x_{i},z) +\sum_{i,j}^{ }\alpha _{i}\alpha _{j}K(x_{i},x_{j})\end{equation}
	Observations in the scoring data set for which $\textup{dist}^{2}(z) > R^{2} $ are designated as outliers.
	
\section{SVDD Bandwidth Selection}
\label{sec:svdd_bandwidth}
Using a kernel function in the SVDD formulation, as outlined in section \ref{df}, is desirable for obtaining a flexible boundary around the training data set. Such a boundary adheres to the essential geometric features of the data and minimizes the misclassification rate. The Gaussian kernel function is the most popular kernel function in SVDD and SVM.  The Gaussian kernel function defined in Eq.\ref{eq:1} has one tuning parameter, the bandwidth parameter $s$. The bandwidth parameter needs to be set before an SVDD model is trained. This section outlines the importance of selecting a good bandwidth value and introduces methods to select such a bandwidth value.

\subsection{Importance of Bandwidth Selection}
The flexible data description is preferred when a data boundary
needs to closely follow the shape of data. The tightness of the boundary is a function of
the number of support vectors. For a Gaussian kernel, it is observed
that if the value of the outlier fraction $f$ is kept constant, the number of
support vectors that are identified by the SVDD algorithm is a function of the Gaussian
bandwidth $s$. At a very low value of $s$, the number of support vectors is
large and approaches the number of observations. As the value of $s$
increases, the number of support vectors is reduced. It is also observed
that the data boundary is extremely wiggly at lower values of $s$.
As $s$ increases, the data boundary becomes less wiggly
and starts to follow the shape of the data.\\

Because SVDD is an unsupervised technique, cross validation cannot be used to determine an appropriate value of $s$. There are several methods for setting an
appropriate kernel bandwidth value. Some of the unsupervised methods include the VAR criterion method\cite{Khazai2012}, the mean criterion method\cite{chau8215749}, the peak criterion method\cite{kakde2017peak,pered8258344}, the method of coefficient of variation (CV) \cite{evangelista2007some}, the method of maximum distance (MD) \cite{khazai2011anomaly}, and the method of distance to the farthest neighbor (DFN) \cite{xiao2014two}. It has been shown on simulated data that the peak criterion method achieves better classification performance than the MD, CV, and DFN methods\cite{kakde2017peak}. The following sections provide more information about the VAR, mean, and peak criterion methods.

\subsection{VAR Criterion Method}
Khazai et al. have proposed a simple SVDD kernel bandwidth selection criterion for hyperspectral data processing: the square root of the sum of the variances of all data variables\cite{Khazai2012}. Given $p$ variables, the selected kernel bandwidth is defined as
\begin{equation}
s = \big(\sum_{j=1}^{p}\sigma^2_j\big)^\frac{1}{2}
\end{equation}
where $\sigma^2_j$ is the variance of the $j_{th}$ variable of the data. 

\subsection{Mean Criterion Method}
The mean criterion \cite{chau8215749} also provides a closed-form expression to obtain the bandwidth value $s$. The mean criterion method uses the fact that when the bandwidth value $s$ is close to $0$ ($s \rightarrow 0^{+}$), the kernel function $k(x_i, x_j)$ that uses any two observations $x_i$ and $x_j$ evaluates to 0 when $i \ne j$ or to 1 when $i = j$. Therefore, when $s$ is close to 0, if the training data set contains $N$ observations, then the kernel matrix of $k(x_i, x_j)$ entries is an identity matrix. Hence, any selected bandwidth value should be large enough to be able to distinguish the kernel matrix from the identity matrix. The mean criterion provides the value of $s$ as
\begin{equation}
\label{eq:mean}
s = \sqrt{\frac{2N\sum_{j=1}^{p}\sigma^2_j}{(N-1)\ln(\frac{N-1}{\delta^2})}} \, 
\end{equation}
where $N$ is the number of training samples, $p$ is the number of dimensions of the training data, $\sigma^2_{i}  (i = 1, 2, ..., p)$ is the data variance in each dimension, and $\delta$ is a tolerance factor that indicates distance from the identity matrix. Larger values of $\delta$ ensure greater distance from the identity matrix. 

The mean criterion method is implemented in the SVDD procedure in SAS$^{\tiny{\textregistered}}$ Visual Data Mining and Machine Learning \cite{institute2017sas}. 

\subsection{Peak Criterion Method}

The peak criterion \cite{kakde2017peak,pered8258344} method requires first solving an SVDD training problem by using different values of bandwidth $s$. It recommends the value of $s$ for which the second derivative of the optimal dual objective function value with respect to $s$ first reaches 0. The experimentation results presented in Ref~\citenum{kakde2017peak, pered8258344} indicate that the peak criterion provides a good $s$ value for obtaining the training data description. 

\subsection{Modified Mean Criterion Method}
Using the peak criterion method to select the proper kernel bandwidth, SVDD usually can obtain a good data boundary that closely follows the training data shape \cite{kakde2017peak,pered8258344}. But the disadvantage of the peak criterion method is that it takes long time to obtain the desired kernel bandwidth because it has to generate the objective function curve by varying the choices of kernel bandwidth, usually a couple of hundred times for a smooth curve.

This paper proposes a new automatic, unsupervised Gaussian kernel bandwidth selection approach, which can perform nearly as well as the peak criterion method while being as time-efficient as the mean criterion method. 

For the kernel bandwidth of mean criterion method (defined in Eq.\ref{eq:mean}), and a specific data set, the variance of the data and the number of training samples $N$ is fixed. So $s$ can be rewritten as a function of $\delta$,
\begin{equation}
\begin{split}
s & = \sqrt{\frac{2N\sum_{j=1}^{p}\sigma^2_j}{N-1}}\sqrt{\frac{1}{\ln(\frac{N-1}{\delta^2})}} \,  \\
  & = \sqrt{\frac{2N\sum_{j=1}^{p}\sigma^2_j}{N-1}} \, \overline{s}(N,\delta) \, 
\end{split}
\end{equation}

where ${\overline{s}(N,{\delta})}$ is a function of the number of observations in the training data set $N$ and the tolerance factor $\delta$,  and is expressed as:\\

\begin{equation} \label{eq:5} 
\overline{s}(N,\delta) = \sqrt{\frac{1}{\ln(\frac{N-1}{\delta^2})}}\,=[\ln(N-1)-2\ln(\delta)]^{-\frac{1}{2}} 
\end{equation}
For a training data set that has a fixed $N$, differentiating with respect to $\delta$ results in the following:
\begin{equation}
\begin{split}
\frac{\partial\overline{s}(N,\delta)}{\partial\delta} & = -\frac{1}{2}[\ln(N-1)-2\ln(\delta)]^{-\frac{3}{2}}(-2)\frac{1}{\delta} \, \\
& = [\ln(N-1)-2\ln(\delta)]^{-\frac{3}{2}}\frac{1}{\delta} \, \\
& = \frac{\overline{s}(N,\delta)^3}{\delta}
\end{split}
\end{equation}
For this paper, experiments were conducted on several data sets that have different numbers of variables $p$ and different numbers of observations $N$. The experiments revealed that the kernel bandwidth value that provides a good classification performance usually happens when $\partial\overline{s}(N,\delta)$ is close to $\partial\delta$. This observation is formalized into the following criterion to select a kernel bandwidth for SVDD:
\begin{equation}
\begin{split}
\frac{\partial\overline{s}{(N,\delta)}}{\partial\delta}& = 1 \\
\end{split}
\end{equation}
This criterion is equivalent to the following:
\begin{equation}\label{eq:6}
\begin{split}
\frac{\overline{s}{(N,\delta)}^3}{\delta} &= 1 \\
\overline{s}{(N,\delta)}^3 & = \delta
\end{split}
\end{equation}

Obtaining the desired kernel bandwidth $s$ with the new selection criterion involves three steps:
\begin{enumerate}
	\item Solve for $\delta$.
	Using Eq. \ref{eq:6} and \ref{eq:5},
	\begin{equation}\label{eq:7}
	\begin{split}
	\delta &= \overline{s}{(N,\delta)}^3 \\
    \delta &= [\ln(N-1)-2\ln(\delta)]^{-\frac{3}{2}} \\
    \end{split}
    \end{equation}	
    \item Use fixed-point iteration \cite{Burden1985} to obtain the value of $\delta$ for a fixed value of $N$ by setting\\
   \begin{equation}\label{eq:9}
    \begin{split}
    \delta{_0} &= 1 \\
    \delta{_{n+1}} &= \overline{s}{(N,\delta{_{n}})}^3 = [\ln(N-1)-2\ln(\delta{_{n}})]^{-\frac{3}{2}} \quad  n=0,1,2...\\
    \end{split}
    \end{equation}
  
    \item Repeat steps 1 and 2 for different values of $N$, where $N$ is the number of observations in the training data set. For a majority of $N$ values, convergence was obtained in three to four iterations. Empirically, it is observed that the value of $\delta$ is approximately polynomial in $\frac{1}{\ln(N-1)}$ with a mean squared error of 7.02E-11 and can be expressed as
    	\begin{equation}
    	   \begin{split}
       \delta &= -0.14818008\phi^4 + 0.284623624\phi^3 - 0.252853808\phi^2 + 0.159059498\phi - 0.001381145 \\
      \quad \text{where } \phi &= \frac{1}{\ln(N-1)}
          \end{split}
    	\end{equation}
   Figure 1 shows the relationship between $\delta$ and $\frac{1}{\ln(N-1)}$. Figure 2 shows the relationship between $\delta$ and $N$. For a given data set that the number of observations $N$ is known, the corresponding $\delta$ can be obtained easily by using the $\delta$ vs. $N$ curve. 
\begin{figure}[h]
	\begin{center}
	\includegraphics[angle=0,origin=c, scale=0.6]{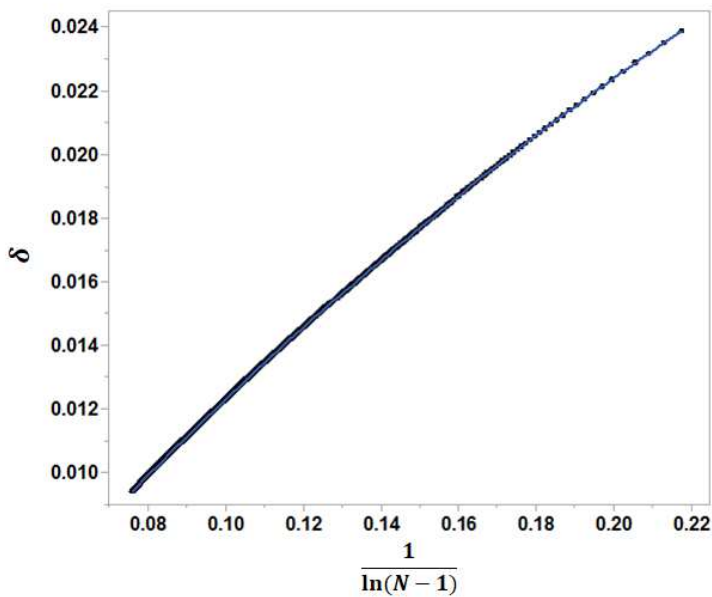}
	\captionof{figure}{Relationship between $\delta$ and $\frac{1}{\ln(N-1)}$}
	\end{center}
\end{figure}
\begin{figure}[h]
	\begin{center}
	\includegraphics[angle=0,origin=c, scale=0.6]{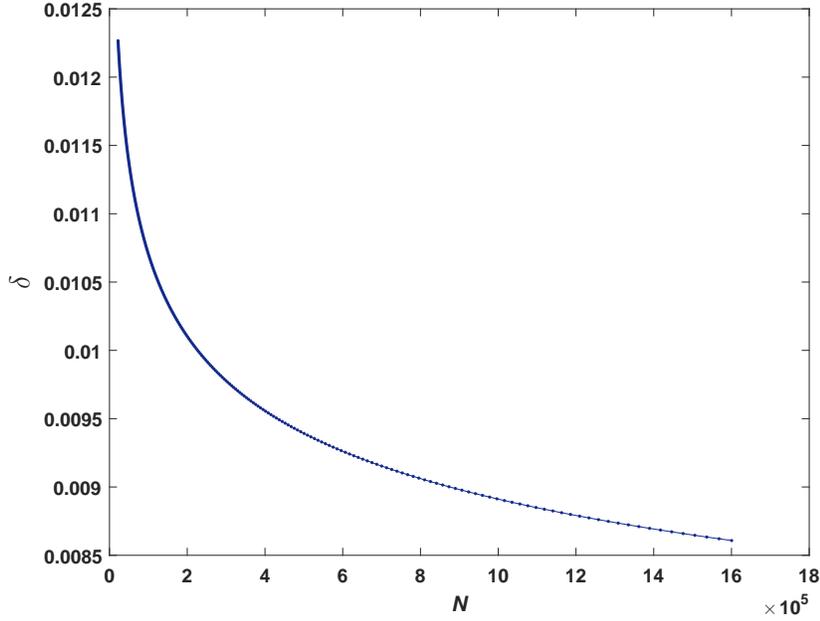}
	\captionof{figure}{Relationship between $\delta$ and $N$}
	\end{center}
\end{figure}
    
	\item After the value of $\delta$ for a particular value of $N$ is obtained, compute the kernel bandwidth $s$ as follows:
	\begin{equation*}
	s = \sqrt{\frac{2N\sum_{j=1}^{p}\sigma^2_j}{(N-1)\ln(\frac{N-1}{\delta^2})}} \,
	\end{equation*}
\end{enumerate}

\section{Data Experiments}
\label{sec:experiments}
\subsection{Data Description}
To evaluate the performance of the new kernel bandwidth selection method, the SVDD classifier was applied to three commonly used hyperspectral data sets: Botswana, Kennedy Space Center (KSC), and Indian Pines\cite{PURR1947}. Table 1 summarizes the main characteristics of these data sets. Table 2 lists all the classes in each data set and the number of ground-truthed samples available for training and testing. 

\begin{table}[ht]
	\caption{Hyperstral Data Sets Summary} 
	\label{tab:Summary}
	\begin{center}       
		\begin{tabular}{|l|c|c|c|} 
			\hline
			\rule[-1ex]{0pt}{3.5ex}  \textbf{Data Set}  & \textbf{Botswana} & \textbf{KSC}  & \textbf{Indian Pines}\\
			\hline
			\rule[-1ex]{0pt}{3.5ex}  Sensor Type  & Hyperion & AVIRIS  & AVIRIS \\
			\hline
			\rule[-1ex]{0pt}{3.5ex}  Spatial Resolution  & 30 m & 18 m  & 20 m\\
			\hline
			\rule[-1ex]{0pt}{3.5ex}  Image Size  & 1476$\times$256 & 512$\times$614  & 145$\times$145 \\
			\hline
			\rule[-1ex]{0pt}{3.5ex}  \# of Spectral Bands & 145 & 176   & 200 \\
			\hline
			\rule[-1ex]{0pt}{3.5ex}  \# of Classes  & 14 & 13  & 16 \\
			\hline		
		\end{tabular}
	\end{center}
\end{table} 
\begin{table}[ht]
	\caption{Hyperspectral Data Set Classes and Number of Samples Available} 
	\begin{center}     
		\scriptsize
		\begin{tabular}{|R{10mm}|p{22mm}|R{15mm}|p{22mm}|R{15mm}|p{22mm}|R{15mm}|}
			\hline
			&
      		\multicolumn{2}{c|}{\textbf{Bostwana}} &
			\multicolumn{2}{c|}{\textbf{KSC}} &
			\multicolumn{2}{c|}{\textbf{Indian Pines}} \\			
			\hline
		     \rule[-1ex]{0pt}{3.5ex} \textbf{Class \#} &  \textbf{Class Name} & \textbf{\# of Samples} & \textbf{Class Name} & \textbf{\# of Samples} & \textbf{Class Name} & \textbf{\# of Samples}  \\
			\hline
			\rule[-1ex]{0pt}{3.5ex} 1 & Water & 270 & Scrub & 761 & Alfalfa & 46  \\
			\hline
			\rule[-1ex]{0pt}{3.5ex} 2 & Hippo Grass & 101 & Willow swamp & 243 & Corn-notill & 1428  \\
			\hline
			\rule[-1ex]{0pt}{3.5ex} 3 & Floodplain grasses 1 & 251 & Cabbage palm hammock & 256 & Corn-mintill & 830  \\
			\hline
			\rule[-1ex]{0pt}{3.5ex} 4 & Floodplain grasses 2 & 215 & Cabbage palm / oak hammock & 252 & Corn & 237  \\
			\hline
			\rule[-1ex]{0pt}{3.5ex} 5 & Reeds & 269 & Slash pine & 161 & Grass-pasture & 483  \\
			\hline
			\rule[-1ex]{0pt}{3.5ex} 6 & Riparian & 269 & Oak/broadleaf hammock & 229 & Grass-trees & 730  \\
			\hline
			\rule[-1ex]{0pt}{3.5ex} 7 & Firescar & 259 & Hardwood swamp & 105 & Grass-pasture-mowed & 28  \\ 
			\hline
			\rule[-1ex]{0pt}{3.5ex} 8 & Island interior & 203 & Spartina marsh & 431 & Hay-windrowed & 478  \\
			\hline
			\rule[-1ex]{0pt}{3.5ex} 9 & Acacia woodlands & 314 & Spartina marsh & 520 & Oats & 20  \\
			\hline
			\rule[-1ex]{0pt}{3.5ex} 10 & Acacia shrublands  & 248 & Cattail marsh & 404 & Soybean-notill & 972  \\
			\hline
			\rule[-1ex]{0pt}{3.5ex} 11 & Acacia grasslands & 305 & Salt marsh & 419 & Soybean-mintill & 2455  \\
			\hline
			\rule[-1ex]{0pt}{3.5ex} 12 & Short mopane  & 181 & Mud flats & 503 & Soybean-clean & 593  \\
			\hline
			\rule[-1ex]{0pt}{3.5ex} 13 & Mixed mopane  & 268 & Water & 927 & Wheat & 205  \\
			\hline
			\rule[-1ex]{0pt}{3.5ex} 14 & Exposed soils & 95 &  &  & Woods & 1265  \\
			\hline
			\rule[-1ex]{0pt}{3.5ex} 15 &  &  &  &  & Building-grass-trees-drives & 386  \\
			\hline
			\rule[-1ex]{0pt}{3.5ex} 16 &  &  &  &  & Stone-steel-towers & 93  \\
			\hline
		\end{tabular}
	\end{center}
\end{table}

\subsection{Evaluation Process}
The evaluation process consists of three steps: data training, data testing, and performance evaluation. The following data preprocessing steps were required before the SVDD approach was applied:
\begin{enumerate}
	\item A special preprocessing step was applied to the KSC data set. Some pixels have saturated values at certain spectral bands; that is, some data values are greater than 65,500 whereas the normal data range is [0, 1244]. These saturated data values were corrected by substituting 0 for them. 
	\item Each data set was normalized with the maximum data value in the set, making the data range always $[0, 1]$\cite{Khazai2012}. 
\end{enumerate}

\subsubsection{Training and Testing}
SVDD is a one-class classifier. In order to solve the multiclass classification problem for hyperspectral data, the same fusion scheme as in Ref.~\citenum{Khazai2012} was used. For each class, an SVDD classifier was trained by using 30\% of the available samples, randomly selected.  The remaining 70\% was reserved for testing. Assuming that there are $M$ classes, each test sample is evaluated against each trained class to obtained its distance $\mbox{dist}_{i}$ to the class's hypersphere center, where $i = 1, 2, ..., M$. A class label is assigned to the test sample on the basis of the following fusion rule\cite{Khazai2012}:
\begin{itemize}
	\item If $\text{dist}_i$ is within the hypersphere radius of only one class, then the label of this class is assigned to the test sample. 
	\item If $\text{dist}_i$ is within the hypersphere radius of more than one class or no classes, the class to be assigned is decided by the following criterion, where $R_i$ is the radius of the hypersphere for class $i$:
		\begin{equation}
 			\operatorname*{arg\,min}_{i=1, 2, ..., M} \frac{\text{dist}_i}{R_i}
		\end{equation}
\end{itemize}

The preceding decision rule is illustrated in Fig.~\ref{fig:fusion}. In this two-class classification example, a test sample $z$'s distance to Class $A$'s hypersphere center,  $\text{dist}_{1}$,  is the same as its distance to Class $B$'s hypersphere center, $\text{dist}_{2}$. Because $R_{A}$ is less than $R_{B}$, the relative distance $\frac{\text{dist}_{1}}{R_{A}}$ is greater than $\frac{\text{dist}_{2}}{R_{B}}$, so the test sample is labeled as Class $B$. 
	\begin{figure} [ht]
		\begin{center}
			\begin{tabular}{c} 
				\includegraphics[height=3cm]{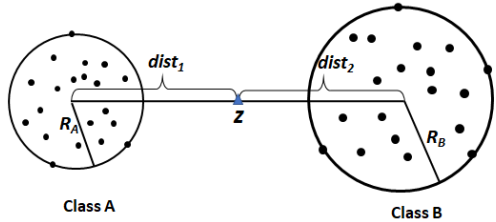}
			\end{tabular}
		\end{center}
		\caption[example] 
		{ \label{fig:fusion} 
			Illustration of the decision rule when a test sample's distances to hypersphere centers of two different classes are equal.}
	\end{figure} 

\subsubsection{Evaluation}
The classification performance was evaluated on four different SVDD kernel bandwidth selection methods that use the VAR criterion\cite{Khazai2012}, the mean criterion\cite{chau8215749}, the peak criterion\cite{kakde2017peak,pered8258344}, and the new modified mean criterion. 

For every data set, the training and testing experiments were carried out five times, each with a different randomly selected subset (30\%) for training and the rest (70\%) for testing. The classification performance was evaluated using the overall accuracy (OA)\cite{Khazai2012}, which is defined as the percentage of pixels that are correctly labeled. 

\subsubsection{Results}
Table 3 through Table 6 show the evaluation results for each hyperspectral data set. Exp$_{1}$ through Exp$_{5}$ represents each experiment, and the last row shows the average overall accuracy of the five experiments. 

From Table 4 (results on the raw KSC data) and Table 5 (results on the corrected KSC data), you can see that the preprocessing step, which replaces the saturated data values with 0, has significantly improved the data classification performance. 

The classification performance results demonstrate that the new modified mean criterion performed uniformly better than other bandwidth selection methods for Botswana, corrected KSC, and Indiana Pine data sets. Because the new method has a closed-form formula of the kernel bandwidth, its time-efficiency is equivalent to that of the VAR and mean criterion methods. The superiority in performance and speed presents a lot of potential for using the new method for other hyperspectral image data processing. 

\begin{table}[!ht]
	\caption{Overall Performance (OA\%) of the Botswana Data set} 
	\begin{center}       
		\begin{tabular}{|l|R{15mm}|R{15mm}|R{15mm}|r|} 
			\hline
			\rule[-1ex]{0pt}{3.5ex}  \textbf{Method}  & \textbf{VAR}  & \textbf{Mean} & \textbf{Peak} & \textbf{Modified Mean} \\
			\hline
			\rule[-1ex]{0pt}{3.5ex}  Exp$_{1}$ & 84.91 & 80.60 & 87.42 & 89.88 \\
            \hline	
			\rule[-1ex]{0pt}{3.5ex}  Exp$_{2}$ & 84.87 & 79.01 & 86.90 & 87.02 \\
			\hline
			\rule[-1ex]{0pt}{3.5ex}  Exp$_{3}$ & 85.00 & 80.91 & 89.09 & 88.91 \\
			\hline
			\rule[-1ex]{0pt}{3.5ex}  Exp$_{4}$ & 84.43 & 81.48 & 88.87 & 86.19 \\
			\hline
			\rule[-1ex]{0pt}{3.5ex}  Exp$_{5}$ & 83.55 & 79.10 & 85.88 & 86.05 \\
			\hline
			\rule[-1ex]{0pt}{3.5ex}  Average & 84.55 & 80.22 & 87.63 & 87.61 \\
			\hline
		\end{tabular}
	\end{center}
\end{table} 

\begin{table}[!ht]
	\caption{Overall Performance (OA\%) of the KSC Data set, Raw Data} 
	\begin{center}       
		\begin{tabular}{|l|R{15mm}|R{15mm}|R{15mm}|r|} 
			\hline
			\rule[-1ex]{0pt}{3.5ex}  \textbf{Method}  & \textbf{VAR}  & \textbf{Mean} & \textbf{Peak} & \textbf{Modified Mean} \\
			\hline
			\rule[-1ex]{0pt}{3.5ex}  Exp$_{1}$ & 46.12 & 49.03 & 49.88 & 49.36 \\
			\hline	
			\rule[-1ex]{0pt}{3.5ex}  Exp$_{2}$ & 35.45 & 33.78 & 28.49 & 33.34 \\
			\hline
			\rule[-1ex]{0pt}{3.5ex}  Exp$_{3}$ & 21.94 & 36.47 & 35.56 & 36.41 \\
			\hline
			\rule[-1ex]{0pt}{3.5ex}  Exp$_{4}$ & 66.03 & 66.41 & 54.13 & 62.22 \\
			\hline
			\rule[-1ex]{0pt}{3.5ex}  Exp$_{5}$ & 58.29 & 60.52 & 82.64 & 60.57 \\
			\hline
			\rule[-1ex]{0pt}{3.5ex}  Average & 45.57 & 49.24 & 50.14 & 48.38 \\
			\hline
		\end{tabular}
	\end{center}
\end{table}

\begin{table}[!ht]
	\caption{Overall Performance (OA\%) of the KSC Data set, Corrected Data} 
	\begin{center}       
		\begin{tabular}{|l|R{15mm}|R{15mm}|R{15mm}|r|} 
			\hline
			\rule[-1ex]{0pt}{3.5ex}  \textbf{Method}  & \textbf{VAR}  & \textbf{Mean} & \textbf{Peak} 
			& \textbf{Modified Mean} \\
			\hline
			\rule[-1ex]{0pt}{3.5ex}  Exp$_{1}$ & 66.03 & 83.58 & 80.42 & 85.00 \\
			\hline	
			\rule[-1ex]{0pt}{3.5ex}  Exp$_{2}$ & 68.08 & 83.14 & 79.35 & 84.10 \\
			\hline
			\rule[-1ex]{0pt}{3.5ex}  Exp$_{3}$ & 66.03 & 84.15 & 79.19 & 86.04 \\
			\hline
			\rule[-1ex]{0pt}{3.5ex}  Exp$_{4}$ & 72.00 & 83.91 & 81.33 & 85.30 \\
			\hline
			\rule[-1ex]{0pt}{3.5ex}  Exp$_{5}$ & 69.92 & 80.89 & 79.52 & 82.75 \\
			\hline
			\rule[-1ex]{0pt}{3.5ex} Average & 68.41 & 83.13 & 79.96 & 84.64 \\
			\hline
		\end{tabular}
	\end{center}
\end{table} 

\begin{table}[!ht]
	\caption{Overall Performance (OA\%) of the Indiana Pine Data set} 
	\begin{center}       
		\begin{tabular}{|l|R{15mm}|R{15mm}|R{15mm}|r|} 
			\hline
			\rule[-1ex]{0pt}{3.5ex}  \textbf{Method}  & \textbf{VAR}  & \textbf{Mean} & \textbf{Peak} 
			& \textbf{Modified Mean} \\
			\hline
			\rule[-1ex]{0pt}{3.5ex}  Exp$_{1}$ & 38.25 & 54.97 & 49.27 & 57.42 \\
			\hline	
			\rule[-1ex]{0pt}{3.5ex}  Exp$_{2}$ & 36.47 & 54.38 & 50.08 & 57.87 \\
			\hline
			\rule[-1ex]{0pt}{3.5ex}  Exp$_{3}$ & 41.78 & 55.35 & 51.89 & 57.26 \\
			\hline
			\rule[-1ex]{0pt}{3.5ex}  Exp$_{4}$ & 33.46 & 53.44 & 47.00 & 56.85 \\
			\hline
			\rule[-1ex]{0pt}{3.5ex}  Exp$_{5}$ & 41.17 & 46.90 & 42.81 & 51.36 \\
			\hline
			\rule[-1ex]{0pt}{3.5ex}  Average & 38.23 & 53.01 & 48.21 & 56.15 \\
			\hline
		\end{tabular}
	\end{center}
\end{table} 

Of the three hyperspectral test data sets---Bostswana, KSC (corrected data), and Indian Pines---the Indian Pine set has the lowest overall accuracy. The classification performance was further analyzed by computing the accuracy of each class and is shown in Table 7. For classes that contain very few labeled samples (Alfalfa, Grass-pasture-mowed, and Oats), there were only $10-15$ training samples per class (which obviously is not enough to characterize the class), and the trained classifier is not able to identify test samples well. The second type of difficulty is in classes that are very similar to each other (for example, Corn-mintill and Core; and Soybean-notill, Soybean-mintill, and Soybean-clean). Given the similar spectral radiance of these materials, misclassification is significant between these classes, and thus has a lower overall accuracy. 

\begin{table}[ht]
	\caption{Accuracy (\%) per Class of Indian Pine Data Set} 
	\begin{center}     
		\scriptsize 
		\begin{tabular}{|r|l|r|r|r|r|r|r|r|}
			\hline
			\rule[-1ex]{0pt}{3.5ex} \textbf{Class} \# &  \textbf{Class Name} & \textbf{\# of Samples} & \textbf{Exp$_{1}$} & \textbf{Exp$_{2}$} & \textbf{Exp$_{3}$} & \textbf{Exp$_{4}$} & \textbf{Exp$_{5}$} & \textbf{Average} \\
			\hline
			\rule[-1ex]{0pt}{3.5ex} 1 & Alfalfa & 46  & 6.25 & 12.50 & 6.25 & 6.25 & 9.38 & 8.13\\
			\hline
			\rule[-1ex]{0pt}{3.5ex} 2 & Corn-notill & 1428 &45.50 & 54.10 & 46.40 & 55.50 & 42.90 & 48.88\\
			\hline
			\rule[-1ex]{0pt}{3.5ex} 3 & Corn-mintill & 830 & 20.31  & 37.52 & 13.77 & 38.04 & 14.46 & 24.82\\
			\hline
			\rule[-1ex]{0pt}{3.5ex} 4 & Corn & 237  & 68.07 & 53.61  & 72.89 & 72.29 & 68.67 & 67.11\\
			\hline
			\rule[-1ex]{0pt}{3.5ex} 5 & Grass-pasture & 483  & 78.99  & 83.43 & 78.99  & 82.84 & 68.34 & 78.52\\
			\hline
			\rule[-1ex]{0pt}{3.5ex} 6 & Grass-trees & 730  & 72.21 & 54.99 & 51.86 & 50.10 & 60.47 & 57.93\\
			\hline
			\rule[-1ex]{0pt}{3.5ex} 7 & Grass-pasture-mowed & 28 & 0 & 0 & 30.00 & 0 & 10.00 & 8.00\\ 
			\hline
			\rule[-1ex]{0pt}{3.5ex} 8 & Hay-windrowed & 478 & 96.72 & 97.61 & 97.91 & 98.21 & 98.21 & 97.73\\
			\hline
			\rule[-1ex]{0pt}{3.5ex} 9 & Oats & 20  & 0  & 0 & 0 & 7.14 & 0 & 1.43 \\
			\hline
			\rule[-1ex]{0pt}{3.5ex} 10 & Soybean-notill & 972  & 32.65 & 38.38  & 30.88 & 27.21 & 26.76 & 31.18\\
			\hline
			\rule[-1ex]{0pt}{3.5ex} 11 & Soybean-mintill & 2455 & 46.57 & 43.42 & 55.41 & 41.85 & 32.89 & 44.03\\
			\hline
			\rule[-1ex]{0pt}{3.5ex} 12 & Soybean-clean & 593  & 78.31 & 75.42 & 76.14 & 68.67 & 81.93 & 76.10\\
			\hline
			\rule[-1ex]{0pt}{3.5ex} 13 & Wheat & 205 & 41.96  & 14.69 & 23.78 & 38.46 & 23.08 & 28.39\\
			\hline
			\rule[-1ex]{0pt}{3.5ex} 14 & Woods & 1265 &  94.46 & 92.77 & 93.56 & 93.33 & 90.17 & 92.86\\
			\hline
			\rule[-1ex]{0pt}{3.5ex} 15 & Building-grass-trees-drives & 386 & 68.52 & 72.96 & 69.26 & 74.81 & 80.37 & 73.19\\
			\hline
			\rule[-1ex]{0pt}{3.5ex} 16 & Stone-steel-towers & 93  & 66.15 & 76.92 & 72.31 & 64.62 & 73.85 & 70.77\\
			\hline
		\end{tabular}
	\end{center}
\end{table}

\section{Conclusion}
This paper proposes a new automatic, unsupervised Gaussian kernel bandwidth selection method for SVDD and applies it to hyperspectral imaging data classification. This method has a closed-form formula for kernel bandwidth calculation. Experiments have shown that the new method outperforms other commonly used SVDD kernel bandwidth selection methods (VAR criterion, mean criterion, and peak criterion) on three benchmark hyperspectral data sets. Experiments with other simulated high-dimensional data also show the robustness of this method when the data dimension increases. Research will be extended to apply the new approach on more high-dimensional data processing and also to look into the physical interpretation of this method.

\acknowledgments 

Authors would like to thank Anne Baxter, Principal Technical Editor at SAS, for her assistance in creating this manuscript. 

\bibliography{report} 
\bibliographystyle{spiebib} 

\end{document}